\documentstyle[prd,aps,floats,epsfig,preprint]{revtex}
\tightenlines

\begin{document}

\title{Heavy Majorana neutrinos at $e^+e^-$ colliders}

\author{G. Cveti\v c$^{a,b}$, C. S. Kim$^{c}$, C. W. Kim$^{d,e}$}

\address{ $^a$Dept.~of Physics, Bielefeld Univ.,
              33501 Bielefeld, Germany}
\address{ $^b$Dept.~of Physics, Dortmund Univ.,
              44221 Dortmund, Germany}
\address{ $^c$Dept.~of Physics, Yonsei Univ., 
              Seoul 120-749, Korea}
\address{ $^{d}$Korea Institute for Advanced Study,
                Seoul 130-012, Korea}
\address{ $^{e}$Dept.~of Physics \& Astronomy, 
                The Johns Hopkins Univ.,
                Baltimore, MD 21218, USA}
%\address{Email addresses:
%cvetic@physik.uni-bielefeld.de;
%kim@cskim.yonsei.ac.kr (http://phya.yonsei.ac.kr/\~{}cskim/);
%cwkim@jhuvms.hcf.jhu.edu}

%\maketitle

\begin{flushright}
BI-TP-98/37, KIAS-P98048, hep-ph/9812525\\
appears in {\em Physical Review Letters\/} (issue of June 7, 1999)
\end{flushright}

\vspace{1.5cm}

\centerline{{\large \bf
Heavy Majorana neutrinos at $e^+e^-$ colliders}}

\vspace{0.8cm}

\centerline{ G.~Cveti\v c}

\centerline{{\small \it 
Dept.~of Physics, Universit\"at Bielefeld,
33501 Bielefeld, Germany;}}
\centerline{{\small \it  
Dept.~of Physics, Universit\"at Dortmund,
44221 Dortmund, Germany}}

\vspace{0.5cm}

\centerline{C.S.~Kim}

\centerline{{\small \it 
Dept.~of Physics, Yonsei University, 
Seoul 120-749, Korea}} 

\vspace{0.5cm}

\centerline{C.W.~Kim}
 
\centerline{{\small \it 
Korea Institute for Advanced Study, Seoul 130-012, Korea;}}
\centerline{{\small \it 
Dept.~of Physics and Astronomy, The Johns Hopkins
University, Baltimore, MD 21218, USA}}

\begin{abstract}
We investigate possibilities for detecting heavy Majorana neutrinos 
($N$'s) in $e^+e^-$ at LEP200 and future Linear Colliders. 
We concentrate on the processes where the pairs of intermediate 
heavy $N$'s produce a clear signal of total lepton number 
violation ($e^+e^- \to NN \to W^+l^-W^+l^{\prime -}$). 
Such a signal is not possible if the heavy neutrinos are of 
Dirac nature. Our approach is general in the sense that the
intermediate $N$'s can be either on-shell or off-shell. Discussion 
of the relative numerical importance of the $s$ and the $t\!+\!u$ 
channels of the $NN$ production is also included.\\
PACS number(s):  14.60.St, 11.30.Fs, 13.10.+q,13.35.Hb
\end{abstract}

%\maketitle

\setcounter{equation}{0}
\newpage

There has been a significant amount of activity in the high energy
physics community towards discerning the
nature of the neutrino sector. A basic question is:
Are neutrinos Dirac or Majorana particles? 
If there are no right-handed currents,
then it is virtually impossible to
discern the nature of the light neutrinos \cite{Kayser}. 
If there are heavy neutrinos
($M\!\sim\!10^2$ GeV), then present and future experiments
offer a realistic prospect of establishing their nature.
The production cross section
of heavy Majorana neutrinos ($N$'s),
mostly via the $e^+e^-$ collisions,
has therefore been investigated
in the past \cite{Aguilaetal}.
Most of these works have been done within
specific (classes of) models;
and it has been assumed that the center-of-mass (CMS) 
energy $\sqrt{s}$
in the process is high enough for the production of on-shell
(OS) heavy $N$'s. The effects of the {\em off-shell\/} (nOS)
$N$'s have been ignored. Moreover, to our knowledge,
various distributions of $N$'s decay products
[$N \to W^{\pm} {\ell}^{\mp} (\to {\rm jets} + {\ell}^{\mp}$)], 
which are produced in the full reaction and
which can actually be detected, have not been investigated in a 
quantitative way. The main reason for this was that
the expressions for invariant amplitudes with two {\em on-shell\/}
$N$'s apparently do not allow a straightforward calculation
of such distributions.
We note that the detection of events for the reactions
$e^+e^- \to NN \to W^{\pm} {\ell}^{\mp}W^{\pm} {\ell}^{\prime \mp}
\to {\rm jets} + {\ell}^{\mp} {\ell}^{\prime \mp}$,
which violate the total lepton number, would be a clear signal
of the Majorana character of the intermediate neutrinos.

We present some results of calculations
for the afore-mentioned reactions. 
We do not restrict ourselves
to any specific (classes of) models. In contrast to
the available literature, our approach allows us to account also 
for the effects of {\em off-shell\/} (nOS) intermediate $N$'s 
on the cross sections $\sigma$. This enables us to investigate
deviations from the previously known $\sigma$'s,
in the ``2OS'' kinematic
region ($\sqrt{s}\!>\!2M\!>\!2M_W$) where
both intermediate $N$'s can, but need not, be on-shell
-- these deviations are termed ``finite width effects.''
Our approach allows us to calculate the $\sigma$'s,
and various distributions, also in the ``1OS'' region
($2M\!>\!\sqrt{s}\!>\!M\!+\!M_W$) where at most one intermediate
$N$ can (but need not) be on-shell, and in the ``nOS'' region
($M\!+\!M_W\!>\!\sqrt{s}$) where both $N$'s
always have to be off-shell.
Our approach makes possible a straightforward calculation
of various distributions of final particles. 
As an illustration, 
we include an angular distribution of the final state leptons 
${\ell}, {\ell}^{\prime}$.

\begin{figure}[htb]
%\setlength{\unitlength}{1.cm}
%\begin{minipage}[t]{9.cm}
\begin{center}
\epsfig{file=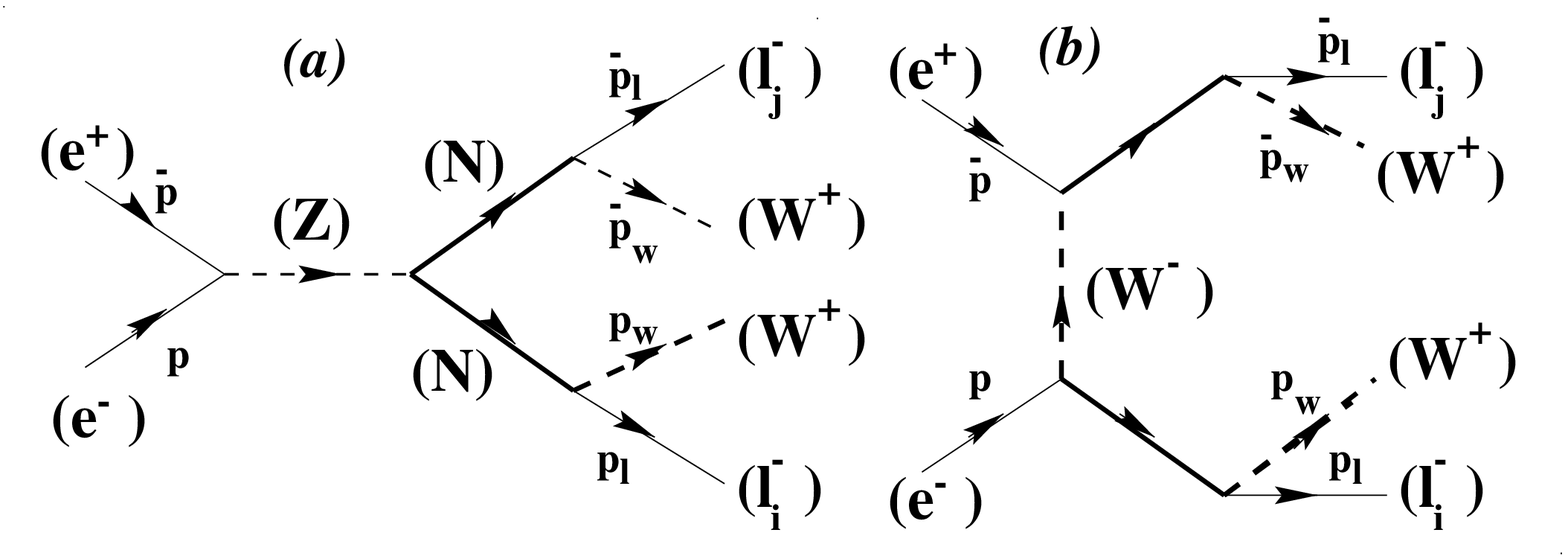,height=5.2cm,width=15.cm}
\end{center}
%\end{minipage}\hfill
%\begin{minipage}[t]{7.cm}
%\epsfig{file=fig1bmaj.eps,height=5.2cm,width=6.7cm}
%\end{minipage}
\vspace{-0.2cm}
\caption{\footnotesize An $s$-channel (a) and a $tu$-channel (b)
diagram for
$e^{\scriptscriptstyle -}(p {\alpha}) 
e^{\scriptscriptstyle +}({\bar p}
{\overline \alpha})\!\to\!N N\!\!\to\!W^
{\scriptscriptstyle +}(p_w {\lambda})
W^{\scriptscriptstyle +}({\bar p}_w {\bar \lambda}) 
{\ell}_i^{\scriptscriptstyle -}(p_{\ell} {\alpha}_{\ell})
{\ell}_j^{\scriptscriptstyle -}({\bar p}_{\ell} {\overline \alpha}_{\ell})$. }
\label{graph}
\end{figure}

We start with rather general Lagrangian densities for the
couplings of $N$ with $Z$, and
for the coupling of $N$ with $W$ and light leptons
${\ell}_j$ (${\ell}_1^-\!=\!e^-$, 
${\ell}_2^-\!=\!\mu^-$, ${\ell}_3\!=\!{\tau}^-$):
\begin{eqnarray}
{\cal {L}}_{NNZ}(x) &=& - \frac{g}{4 \cos \theta_W} A_{NZ} {\overline N}(x)
{\gamma}^{\mu} {\gamma}_5 N(x) Z_{\mu} \ ,
\label{NNZ}
\\
{\cal {L}}_{N{\ell}W}(x) &=& - \sum_{j=1}^3
\frac{ g B_L^{(j)}}{ 2 \sqrt{2} } {\bar \ell}_j(x) 
{\gamma}^{\mu} {\gamma}_{\scriptscriptstyle -} N(x) 
W_{\mu}^{\scriptscriptstyle -}\!+\!{\rm h.c.},
\label{NellW}
\end{eqnarray}
where ${\gamma}_{\scriptscriptstyle -}\!=\!(1\!-\!{\gamma}_5)$;
$B_L^{(j)}$'s are, at first, free parameters;
$g$ and $\theta_W$ are the standard $SU(2)_L$ gauge coupling
parameter and the Weinberg angle, respectively. The vector part 
is absent in (\ref{NNZ}) because $N$'s
are Majorana. The right-handed parts were neglected in (\ref{NellW}).
The other relevant coupling is
$e^{\scriptscriptstyle +}e^{\scriptscriptstyle -}Z$ 
which we consider to be the one of the Standard Model (SM). 
We also set $A_{NZ}\!=\!-1$, i.e., by
replacing, in the SM density for $\nu \nu Z$,
the massless Dirac neutrino $\nu$ by the (heavy) Majorana
neutrino $N$. These choices would suggest that
the considered $N$ is made up primarily of a sequential 
neutrino with the standard
$SU(2)_L\!\times\!U(1)_Y$ assignments. However, these choices
may also represent an approximation to other scenarios
(cf.~\cite{Aguilaetal}, \cite{LL}, \cite{constr}).
Further, parameters $B_L^{(j)}$ in (\ref{NellW})
will affect the final results only via the combinations
\begin{equation}
H1 \equiv |B_L^{(1)}|^2, \quad 
H \equiv \sum_{j=1}^3 |B_L^{(j)}|^2 \ .
\label{H1H}
\end{equation}
We restrict ourselves to the afore-mentioned reactions
$e^+e^- \to NN \to W^{\pm} {\ell}_i^{\mp}W^{\pm} {\ell}_j^{\mp}
(\to {\rm jets}\!+\!{\ell}_i^{\mp} {\ell}_j^{\mp})$
with light leptons (${\ell}_1\!=\!e$, ${\ell}_2\!=\!{\mu}$,
${\ell}_3\!=\!{\tau}$).
They involve the $s$
and the $t\!+\!u$ (shortly: $tu$) channel --
cf.~Fig.~\ref{graph}.

For the calculation of the invariant amplitude
${\cal {M}}_{fi}$ (shortly: ${\cal {M}} \cite{Peskinetal}$)
for various channels, we used the 4-component
spinors $u^{(\alpha)}(q)\!\equiv\!u(q \alpha)$ 
and $v^{(\alpha)}(q)\!\equiv\!v(q \alpha)$ as defined in \cite{IZ},
but with the normalization convention as given
in \cite{Peskinetal}. For the quark
spin components we use the notation:
${\widetilde{\alpha}}\!=\!1, 2 \Leftrightarrow \alpha\!=\!2, 1$.
In the $s$-channel, the resulting amplitude is
\begin{eqnarray}
\lefteqn{
{\rm i} {\cal M}^{(s)} = \frac{ 4 M A^{(s)} } 
{ \left[ s\!-\!M_Z^2\!+\!{\rm i} \Gamma_Z M_Z \right] } 
(-1)^{ {\overline \alpha}_{\ell} }   
\left[ {\bar v}({\bar p} {\overline \alpha}) \gamma_{\mu}
\left( A^{(e)}_{\scriptscriptstyle +}\!-\!A^
{(e)}_{\scriptscriptstyle -} \gamma_5 \right) u(p \alpha)
\right] \times
}
\nonumber\\
&& {\Big \lbrace}
P_N(p_{\ell} p_w) P_N({\bar p}_{\ell} {\bar p}_w) 
{\bar u}(p_{\ell} {\alpha}_{\ell}) 
{\tilde {\cal C}}^{\mu}(p_w {\bar p}_w p_{\ell} {\bar p}_{\ell})
(1\!+\!{\gamma}_5) v({\bar p}_{\ell} {\widetilde {\overline \alpha}}_{\ell} 
)\!+\!(p_w\!\leftrightarrow\!{\bar p}_w)
{\Big \rbrace} \ .
\label{sAmpl1}
\end{eqnarray}
Here, we use notations of Fig.~\ref{graph};
$s\!=\!(p\!+\!{\bar p})^2$; $M_Z$ and $\Gamma_Z$ are the mass and
the total decay width of $Z$;
$A^{(e)}_{\scriptscriptstyle +}$ and 
$A^{(e)}_{\scriptscriptstyle -}$ are the vector and axial-vector
coupling parameters of the 
$e^{\scriptscriptstyle +}e^{\scriptscriptstyle -}Z$ coupling of the SM,
respectively ($A^{(e)}_{\scriptscriptstyle +}\!=\!4 \sin^2 \theta_W\!-\!1$,
$A^{(e)}_{\scriptscriptstyle -}\!=\!-1$). ${\tilde {\cal C}}^{\mu}$'s are:
\begin{equation}
{\tilde {\cal C}}^{\mu}(p_w {\bar p}_w p_{\ell} {\bar p}_{\ell}) 
= {\varepsilon \llap /}  
\left[ ( {p \llap /}_{\ell}\!+\!{p \llap /}_w ){\gamma}^{\mu}
+ {\gamma}^{\mu} ( {{\bar p} \llap /}_{\ell}\!+\!{{\bar p} \llap /}_w )
\right] { {\bar \varepsilon} \llap /} \ ,
\label{tildeC}
\end{equation}
where ${q \llap /}\!\equiv\!q_{\nu} {\gamma}^{\nu}$;
$\varepsilon_{\nu}\!\equiv\!{\varepsilon}_{\nu}^{(\lambda)}(p_w)$
and ${\bar \varepsilon}_{\nu}\!\equiv\!{\varepsilon}_
{\nu}^{(\bar \lambda)}({\bar p}_w)$ are the real
polarization vectors \cite{IZ} of the final $W$'s, with polarizations 
$\lambda,{\bar \lambda}\!=\!1,2,3$. 
$M$ is the mass of $N$'s; $A^{(s)}$ is:
\begin{equation}
A^{(s)} = g^2 B_L^{(i)} B_L^{(j)} A_{NZ} {\rm i} \lambda_M/
(128 \cos^2 \theta_W) \ ,
\label{As}
\end{equation}
where ${\lambda}_M$ is the phase factor in the
Fourier decomposition of the Majorana field $N(x)$
(cf.~\cite{Kayseretal}; $|\lambda_M|^2\!=\!1$). $P_N$
in (\ref{sAmpl1}) is the (scalar) denominator of the
propagator of $N$ 
\begin{equation}
P_N(p_{\ell}p_w) =  
1/[ (p_{\ell}\!+\!p_w)^2\!-\!M^2\!+\!{\rm i} M {\Gamma}_N ] \ ,
\label{PN}
\end{equation}
where ${\Gamma}_N$ is the total decay width of $N$.

All the combinations of attaching the four
final particles to the two $N$'s
are accounted for in amplitude (\ref{sAmpl1}).
Further, (\ref{sAmpl1}) had originally four instead of
two terms in the curly brackets; however, two of them were reduced
to the other two, by using the general identities
\begin{equation}
- {\rm i} {\gamma}^2 u(q {\alpha})^{\ast}\!=\!(-1)^{\alpha} 
v(q {\widetilde {\alpha}}), \
- {\rm i} {\gamma}^2 v(q {\alpha})^{\ast}\!=\!(-1)^{\widetilde {\alpha}}
u(q {\widetilde {\alpha}}) ,
\label{CPtransf}
\end{equation}
and $({a \llap /} {b \llap /} )^T\!=\!- {\gamma}^0 {\gamma}^2
({b \llap /}{a \llap /}) {\gamma}^2 {\gamma}^0$,
where the Dirac basis and the conventions of \cite{IZ}
are used for ${\gamma}^{\mu}$'s.
Using (\ref{CPtransf}), we can further
rewrite (\ref{sAmpl1}) into a form involving
$v(p_{\ell} {\widetilde {\alpha}}_{\ell})$ and
${\bar u}({\bar p}_{\ell} {\overline \alpha}_{\ell})$ instead of
${\bar u}(p_{\ell} {\alpha}_{\ell})$ and
$v({\bar p}_{\ell} {\widetilde {\overline \alpha}}_{\ell} )$.
The complex conjugate ({\em c.c.\/}) of this alternative form 
{\em and\/} of the form
(\ref{sAmpl1}) are needed to calculate later the
$s$-$tu$ interference term of the full $\langle |{\cal M}|^2 \rangle$,
where $\langle \ldots \rangle$ stands for summation over the 
final and average over the initial polarizations.
For the $s$-$s$ part of $\langle |{\cal M}|^2 \rangle$, 
the form (\ref{sAmpl1}) is used.

The $tu$-channel amplitude ${\cal M}^{(tu)}$ turns out to be
\begin{eqnarray}
\lefteqn{
{\rm i} {\cal M}^{(tu)} = 
\frac{4 M A^{(tu)} P_N(p_{\ell}p_w) P_N({\bar p}_{\ell}{\bar p}_w) }
{ [ (p\!-\!p_{\ell}\!-\!p_w)^2\!-\!M_W^2\!+\!{\rm i} {\Gamma}_W M_W ] }
(-1)^{{\overline \alpha}_{\ell}} 
\times 
}
\nonumber\\
&&
{\Big \lbrace}\!\left[ {\bar u}(p_{\ell} {\alpha}_{\ell})
{\tilde {\cal A}}(p_w p_{\ell})
{\gamma}^{\nu} {\gamma}_{\scriptscriptstyle -} u(p {\alpha}) \right] 
\left[ {\bar v}({\bar p} {\overline \alpha})
{\gamma}_{\nu} {{\bar \varepsilon} \llap /} {\gamma}_{\scriptscriptstyle +}
v({\bar p}_{\ell} {\widetilde {\overline \alpha}}_{\ell}) \right] +
\nonumber\\
&&+ \frac{ (p_{\ell}\!+\!p_w)^2 }{M_W^2} 
\left[{\bar u}(p_{\ell} {\alpha}_{\ell}) {\varepsilon \llap /}
{\gamma}_{\scriptscriptstyle -} u(p {\alpha}) \right]
\left[{\bar v}({\bar p} {\overline \alpha}) 
{\cal A}({\bar p}_w {\bar p}_{\ell})
{\gamma}_{\scriptscriptstyle +}
v({\bar p}_{\ell} {\widetilde {\overline \alpha}}_{\ell}) 
\right]\!{\Big \rbrace} \ + \  \ldots ,
\label{tuAmpl1}
\end{eqnarray}
where ${\tilde {\cal A}}(p_w p_{\ell})\!=\!( {\varepsilon \llap /} 
{p \llap /}_w\!+\!2 p_{\ell} \cdot {\varepsilon} )$,
${\cal A}({\bar p}_w {\bar p}_{\ell})\!=\!(
{{\bar p} \llap /}_w {{\bar \varepsilon} \llap /}\!+\!2 
{\bar p}_{\ell} \cdot {\bar \varepsilon} )$, 
${\gamma}_{\scriptscriptstyle \pm}\!=\!(1\!\pm\!{\gamma}_5)$.
The dots at the end of (\ref{tuAmpl1}) stand for three analogous terms,
obtained from the above explicit expression by replacements:
I.~$(p_w,{\varepsilon})\!\leftrightarrow\!({\bar p}_w,
{\bar \varepsilon})$; 
II.~$(p_{\ell},{\alpha}_{\ell})\!\leftrightarrow\!({\bar p}_{\ell},
{\overline \alpha}_{\ell})$ and overall factor $(-1)$;
III.~combined replacements I. and II.
$A^{(tu)}$ in (\ref{tuAmpl1}) is
\begin{equation}
A^{(tu)} = g^4 |B_L^{(1)}|^2 B_L^{(i)} B_L^{(j)} {\rm i} \lambda_M/64  \ .
\label{Atu}
\end{equation}
As in the $s$-channel case,
we can reexpress any of the terms
in ${\cal M}^{(tu)}$ in alternative forms, by applying
transformations (\ref{CPtransf}) -- e.g., if we want to
use, in scalar expressions in square brackets of (\ref{tuAmpl1}), 
$u({\bar p} {\widetilde {\overline \alpha}})$ and
${\bar u}({\bar p}_{\ell} {\overline \alpha}_{\ell})$ instead of
${\bar v}({\bar p} {\overline \alpha})$ and
$v({\bar p}_{\ell} {\widetilde {\overline \alpha}})$. 
Such transformations are convenient when we calculate
$\langle |{\cal M}|^2 \rangle\!\equiv\!\langle 
|{\cal M}^{(s)}\!+\!{\cal M}^{(tu)}|^2 \rangle$.
Then we can always end up with traces 
involving $u(q,{\beta}) {\bar u}(q,{\beta})\!=\!{q \llap /}$
and/or $v(q,{\beta}) {\bar v}(q,{\beta})\!=\!{q \llap /}$
($q\!=\!p,{\bar p},\ldots$; 
${\beta}\!=\!{\alpha},{\overline \alpha},\ldots\!=\!1,2$).

The integrand $\langle |{\cal M}|^2 \rangle$ is long --
the $s$-$tu$ and (even more so) $tu$-$tu$ parts 
extend over tens of pages when printed out. 
Numerical integration of $\langle |{\cal M}|^2 \rangle$
over (parts of) the final phase space leads to the cross sections.
This general (nOS) program, as mentioned, accounts for
the effects of off-shell and on-shell $N$'s.

The input were values of $\sqrt{s}$, $M$,
$H1$ and $H$ [cf.~Eqs.~(\ref{NellW})-(\ref{H1H})].
$H1$ measures the $eWN$ coupling and affects the $tu$-amplitude
($\propto\!H1$). $H$ affects the total 
$\langle |{\cal M}|^2 \rangle$
which is then {\em formally\/} $\propto\!H^2$ (if $H1$ is
fixed). In $\langle |{\cal M}|^2 \rangle$,
we average over the initial (${\alpha},{\overline \alpha}$) ,
and sum over the final polarizations (${\lambda}, {\bar \lambda}$;
${\alpha}_{\ell}, {\overline \alpha}_{\ell}$)
{\em and\/} over the flavors ($i,j\!=\!1,2,3$) of 
the two final light leptons.
In the general (nOS) expression, an additional factor $1/4$
is included in $\langle | {\cal M} |^2 \rangle$ to
avoid double-counting of the two $W^+$'s and of the final leptons
when integrating over the phase space.

By the same described methods we also
calculated ${\cal M}$ and $\langle |{\cal {M}}|^2 \rangle$
when one $N$, or both $N$'s, are explicitly put on-shell 
(1OS, 2OS expressions, respectively). 
The 1OS $\langle |{\cal {M}}|^2 \rangle$, for the sum of reactions
$e^+e^- \to N N_{\rm OS} \to W^{\scriptscriptstyle +} 
{\ell}_i^{\scriptscriptstyle -}  N_{\rm OS}$
($i\!=\!1,2,3$),
was then multiplied by the branching ratio $Br$
for the sum of the decay modes
$N_{\rm OS} \to W^{\scriptscriptstyle +} 
{\ell}_j^{\scriptscriptstyle -}$ ($j\!=\!1,2,3$);
the 2OS $\langle |{\cal {M}}|^2 \rangle$ for 
$e^+e^- \to N_{\rm OS} N_{\rm OS}$ was multiplied by $(Br)^2$.

${\Gamma}_N$, appearing in (\ref{PN}),
was determined at the tree level, assuming that
the only (dominant) decay modes are $N\!\to\!W^{\pm} {\ell}_j^{\mp}$
($j\!=\!1,2,3$) [$\Rightarrow\!{\Gamma}_N\!\propto\!H$].
Then $Br\!=\!1/2$. 

\begin{figure}[htb]
\setlength{\unitlength}{1.cm}
%\begin{minipage}[t]{8.cm}
\begin{center}
\epsfig{file=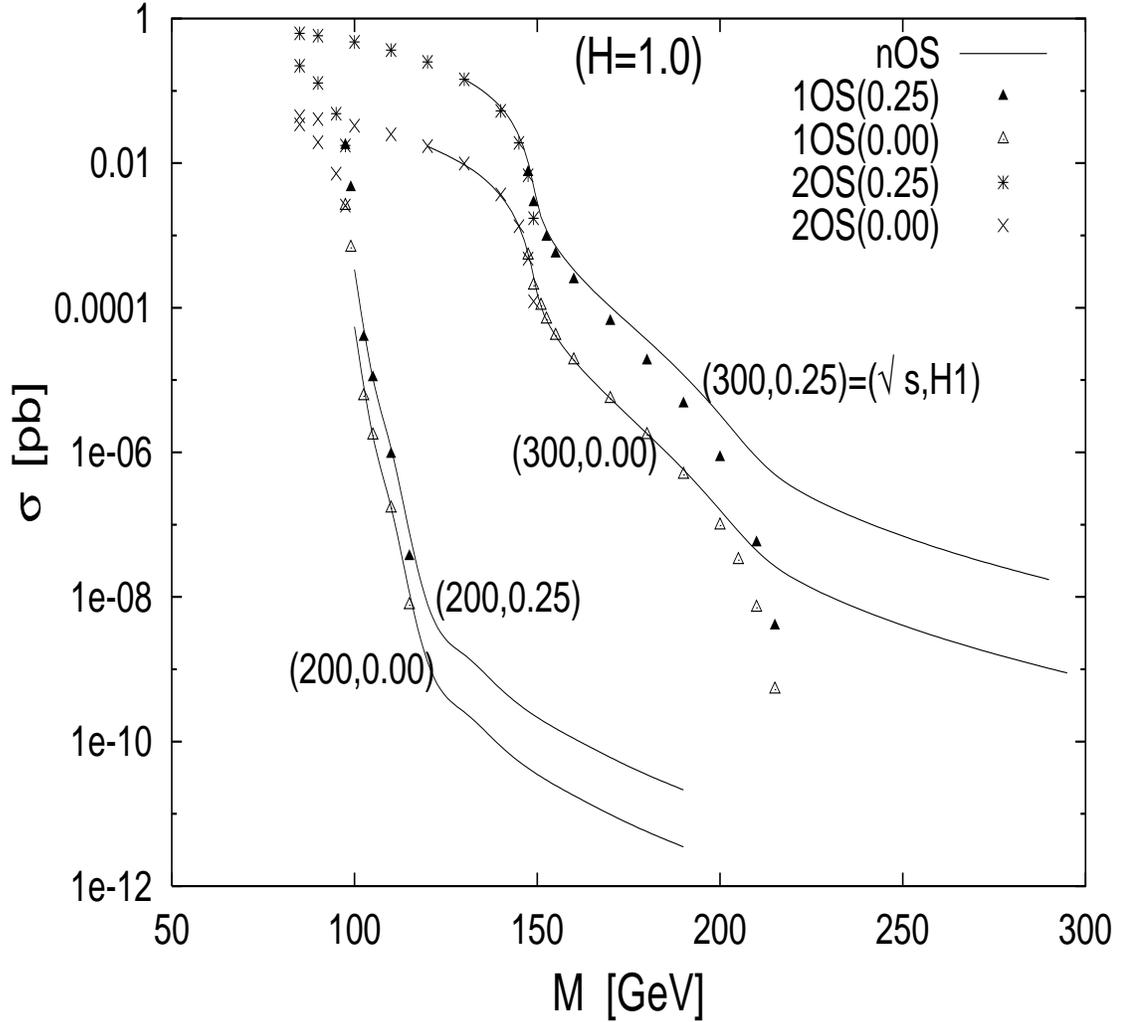,height=14.4cm,width=15.4cm}
\end{center}
%\end{minipage}\hfill
%\begin{minipage}[t]{8.cm}
%\epsfig{file=fig300OS2gen.eps,height=7.2cm,width=7.7cm}
%\end{minipage}
\vspace{-0.4cm}
\caption{\footnotesize 
Sum of cross sections for
$e^+e^-\!\to\!N N\!\to\!W^+ W^+ {\ell}_i^- {\ell}_j^-$
(${\ell}_1\!=\!e$, ${\ell}_2\!=\!{\mu}$, 
${\ell}_3\!=\!{\tau}$), as function of neutrino mass $M$,
for $\sqrt{s}\!=\!200, 300$ GeV and the $tu$-strengths
$H1\!=\!0.0,0.25$.
Full lines are results of the general (nOS) program.
Triangles and crosses are results of the 1OS and 2OS
programs, respectively.}
%Fig.~(b) is enlarged version for the 2OS kin.~regions.}
\label{FigMfun}
\end{figure}

\begin{figure}[htb]
\setlength{\unitlength}{1.cm}
%\begin{minipage}[t]{8.cm}
\begin{center}
\epsfig{file=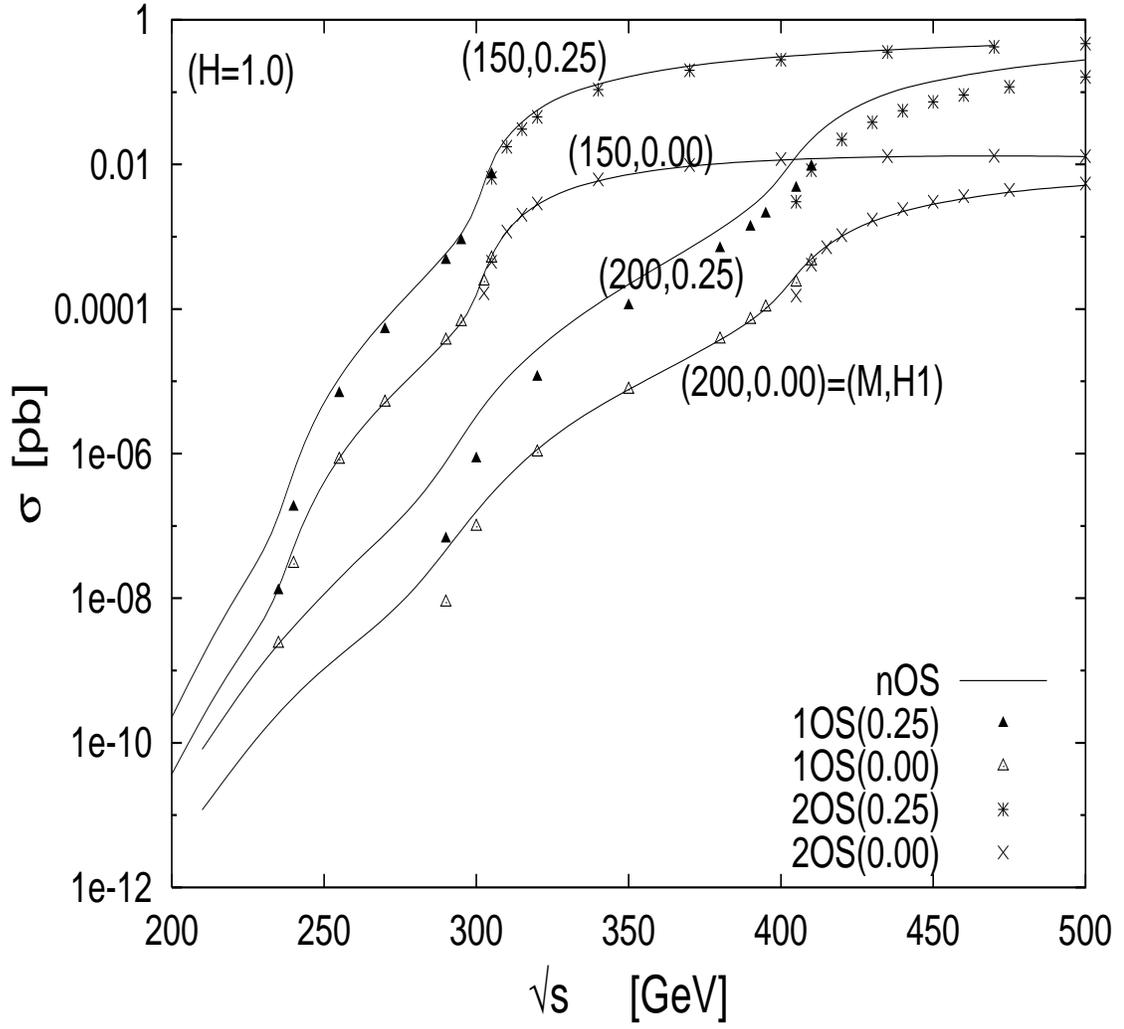,height=14.4cm,width=15.4cm}
\end{center}
%\end{minipage}\hfill
%\begin{minipage}[t]{8.cm}
%\epsfig{file=figM200OS2gen.eps,height=7.2cm,width=7.7cm}
%\end{minipage}
\vspace{-0.4cm}
\caption{\footnotesize Sum of the cross sections for
the mentioned reactions, as function of the CMS
energy $\sqrt{s}$, at fixed $M\!=\!150,200$ GeV. 
Again $H1\!=\!0.0, 0.25$. 
Results of various programs (nOS, 1OS, 2OS) are displayed
in the kinematic regions where they are applicable.}
%Fig.~(b) is enlarged version for the 2OS kin.~regions.
\label{Figrsfun}
\end{figure}

Numerical calculations were
performed in various kinematic regions (nOS, 1OS, 2OS
regions) with the general (nOS) expression [cf.~(\ref{PN})].
In the 1OS and 2OS regions, the 1OS expression was also used.
In the 2OS region, the 2OS expression was also used.
Results are depicted
in Figs.~\ref{FigMfun}-\ref{Figrsfun}. The ${\Gamma}_N$-parameter $H$
was set $H\!=\!1$ in all these Figures.

Figure~\ref{FigMfun} shows the $M$-dependence of the
cross section ${\sigma}$, at fixed $\sqrt{s}$.
The difference between the results of the general
(nOS) and the 2OS program, for $\sqrt{s}\!=\!300$ GeV,
is less then ten percent over most of the 2OS kinematic region
($M_W\!<\!M\!<\!\sqrt{s}/2$), except near the
threshold $M\!\approx\!\sqrt{s}/2$ where the results
of the nOS program are significantly higher.
The difference between the results of the 1OS and 2OS
programs is less than three percent
in most of the 2OS kinematic region.
However, in the 1OS region 
($\sqrt{s}/2\!<\!M\!<\sqrt{s}\!-\!M_W$),
the results of the 1OS program are usually
by several factors lower than those of the
general nOS program, except
near the threshold $M\!\approx\!\sqrt{s}/2$
where they differ only little. All these differences are in general
less pronounced when the $tu$-channel is excluded ($H1\!=\!0.0$). 
For the chosen $tu$-strength value $H1\!=\!0.25$,
the contributions of the $tu$-$tu$ channel are at least by
one order of magnitude larger than those of
the $s$-$s$ channel. Each curve
has two slope increases:
at $M\!\approx\!\sqrt{s}/2$ and at $M\!\approx\!\sqrt{s}\!-\!M_W$
(onset of the 2OS, 1OS kinematic region, respectively). If we take the
integrated luminosity at LEP200 ($\sqrt{s}\!=\!200$ GeV)
to be $500 \ {\rm pb}^{-1}$, Fig.~\ref{FigMfun} implies that
the maximal number of events cannot exceed
$17$ and $112$ if $H1\!=\!0.0, 0.25$, respectively and
we assume $M\!>\!85$ GeV.

In Fig.~\ref{Figrsfun} we show the $\sqrt{s}$-dependence of 
${\sigma}$, at fixed $M$.
Most of the remarks about
Fig.~\ref{FigMfun} apply also to Fig.~\ref{Figrsfun}.
The differences between the results of the nOS and 2OS
programs (``finite width effects'') are very significant when
$M\!=\!200$ GeV (and $H1\!=\!0.25$, $H\!=\!1.0$), because
${\Gamma}_N\!=\!4.8$ GeV is large then
(${\Gamma}_N/M\!\approx\!2.5 \%$; for $M\!=\!150$ GeV:
${\Gamma}_N/M\!\approx\!1.1 \%$).

\begin{figure}[htb]
\begin{center}
\epsfig{file=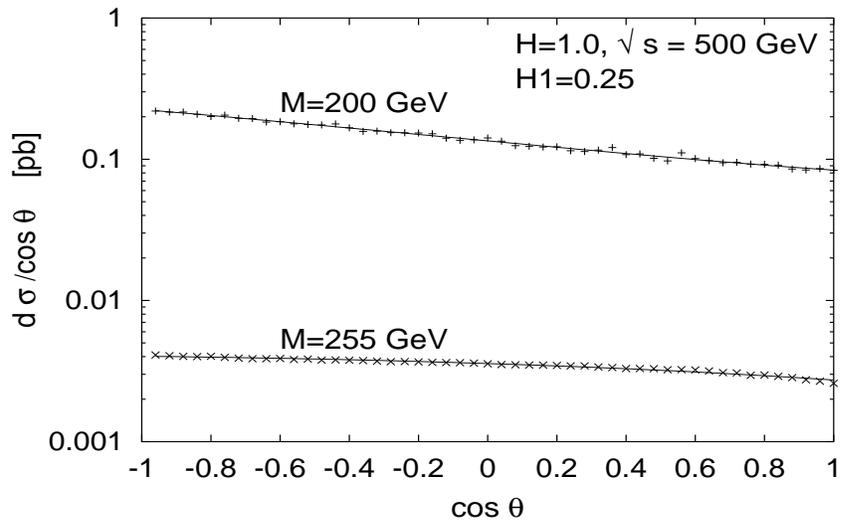,height=7.4cm,width=12.cm}
\end{center}
\vspace{-0.3cm}
\caption{\footnotesize $d {\sigma}/d \cos \theta$,
where ${\theta}$ is the CMS angle between ${\ell}_i, {\ell}_j$.
Input values ($H, H1, \sqrt{s}, M$) are denoted.
The points are results
of calculations, and the curves are parabolas (in $\cos \theta$)
fitted to the points with equal weights. Fluctuations are due to
the limited statistics of the Monte Carlo integration.}
\label{Figd}
\end{figure}

Our general nOS program 
can be applied also to calculation of various 
distributions in the process.
As an illustration, we present in Fig.~\ref{Figd}
an angular distribution of the final leptons ${\ell}^-$, 
${\ell}^{\prime -}$.
The corresponding total cross sections are
${\sigma}\!=\!0.280$ pb, $0.007$ pb, for 
$M\!=\!200, 255$ GeV, and the kinematic regions are 2OS, 1OS, 
respectively.
If linear colliders at $\sqrt{s}\!=\!500$ GeV achieve
the integrated luminosity of
$10^4 \ {\rm pb}^{-1}$, and 
most of the final states $W^+W^+{\ell}_i^-{\ell}_j^-$
can be identified, then these ${\sigma}$'s will correspond
to 2800 and 70 events, respectively.

In the 2OS kinematic region,
the $\sigma$'s
and distributions as given numerically by the general (nOS) program
depend on $H$ weakly. 
Parameter $H$ (note: ${\Gamma}_N\!\propto\!H$) is 
responsible in the 2OS region for the
deviation of the full ${\sigma}$
from the pure 2OS ${\sigma}$. 
In the 1OS region,
$H$-dependence of the full ${\sigma}$ becomes quite strong
(approximately $\sigma\!\propto\!H$), and in the nOS region 
even more so ($\propto$$H^2$).
In Figs.~\ref{FigMfun}-\ref{Figd}, we chose $H\!=\!1$.

In large classes of models, in which heavy neutrinos
are sequential or have exotic $SU(2)\!\times\!U(1)$
assignments, $H1$ and $H$ of (\ref{H1H}) are severely
restricted by available experimental data (LEP and low-energy
data) \cite{LL,constr}: $H1\!<\!0.016$, $H\!<\!0.122$.
In principle, in certain models these restrictions can be avoided.
Taking into account the restriction $H1\!<\!0.016$,
the influence of the $tu$-channel is much weaker
than for $H1\!=\!0.25$, but
the $s$-$tu$ interference term still increases 
the cross section significantly above the pure
$s$-$s$ contributions (cf.~Table \ref{tabl1}).
The restriction $H\!<\!0.122$ would imply that the displayed 
off-shell (``finite width'') effects in the 
2OS region would be weaker than in the $H\!=\!1$ case,
and that the displayed $\sigma$'s in the 1OS and nOS regions
would be reduced, approximately by factors $H$ and
$H^2$, respectively.
The numbers in Table \ref{tabl1} were obtained from the 
general (nOS) program, except in the case of
$M\!=\!85$-$95$ GeV when the 2OS program was used.
%(The reason: the nOS program has numerical
%instabilities in this case, due to a very narrow
%${\Gamma}_N$ (${\Gamma}_N/M\!\approx\!10^{-3}$, when $H\!=\!1.0$)
%and a very restricted final phase space.)

To summarize, we calculated cross sections for
$e^+e^- \to N N \to W^+ {\ell}^- W^+ {\ell}^{\prime -}$
where $N$'s are Majorana neutrinos
(mass $M\!\sim\!10^2$ GeV) 
and ${\ell}, {\ell}^{\prime}$ are light
leptons $e, \mu, \tau$. In contrast to
the calculations available so far, we included the effects of
the {\em off-shell\/} intermediate $N$'s. 
These effects were significant even when
$\sqrt{s}\!>2M$ ($>2 M_W$).
They are more pronounced when 
the $t\!+\!u$ (shortly: $tu$) channel contributions are significant.
The number of reaction events at LEP200
($\sqrt{s}\!=\!200$ GeV; integrated luminosity
$\approx$$500 \ {\rm pb}^{-1}$)
would be low ($<$$20$, for $M\!>\!85$ GeV)
if the strength of the $tu$-channel is
restricted by available experiments {\em and\/} by
confining ourselves to certain classes of models
where the heavy neutrinos are either sequential or
have exotic $SU(2)\!\times\!U(1)$ assignments
(i.e., when $H1\!<\!0.016$). 
Numbers of such events in general significantly
increase at linear colliders ($\sqrt{s}\!=\!500$ GeV; 
integrated luminosity $10^4 \ {\rm pb}^{-1}$), and may be significant
even when $2M\!>\!\sqrt{s}$.
Further, our
approach allows us to calculate numerically
various distributions of the final particles. 

We ignored the questions connected with the
experimental difficulties of detecting the discussed
process unambiguously. In particular, there are
problems connected with identification of the
(on-shell) $W$'s and ${\tau}$'s. 
Further, we ignored the possibility $M\!<\!M_W$
(GC thanks Amitava Datta on that point) --
however, additional
problems arise in the identification of the process
since the two $W^+$'s are then intermediate off-shell.

GC thanks A.~Datta for helpful discussions
and D. Schildknecht for financial support via the BMBF
project no.~332-4005-05-7BI92P toward the end
of the work. CSK thanks W.~Buchm\"uller for helpful
discussions and acknowledges the
financial support of Korean Research Foundation made in the
program year 1997.

\clearpage
\newpage

\begin{table}[htb]
\begin{center}
\begin{tabular}{ c c | l l}
$\sqrt{s}$ [GeV] & $M$ [GeV] & ${\sigma}$ [pb] & increase \\
\hline \hline
500 & 200  & $0.52 \cdot 10^{-2} $ \ ($0.77 \cdot 10^{-2}$) &
$49 \%$ \\
500 & 200  & [$0.53 \cdot 10^{-2}$ \ ($0.75 \cdot 10^{-2}$)] &
[$43 \%$] \\
\hline
500 & 255  & $0.86 \cdot 10^{-4}$ \ ($1.41 \cdot 10^{-4}$) &
$65 \%$ \\
500 & 255  & [$1.14 \cdot 10^{-5}$ \ ($1.88 \cdot 10^{-5}$)] &
[$65 \%$] \\
\hline \hline
300 & 145  & $1.33 \cdot 10^{-3}$ \ ($1.69 \cdot 10^{-3}$) &
$27 \%$ \\
\hline
300 & 155  & $0.41 \cdot 10^{-4}$ \ ($0.52 \cdot 10^{-4}$) &
$27 \%$ \\
\hline \hline
200 & 85  & $0.34 \cdot 10^{-1}$ \ ($0.40 \cdot 10^{-1}$) &
$17 \%$ \\
\hline
200 & 95  & $0.72 \cdot 10^{-2}$ \ ($0.84 \cdot 10^{-3}$) &
$18 \%$ \\
\hline
200 & 105  & $1.70 \cdot 10^{-5}$ \ ($1.96 \cdot 10^{-5}$) &
$15 \%$ \\
%\hline 
\end{tabular}
\end{center}
\vspace{-0.1cm}
\caption {\footnotesize Values of cross sections ${\sigma}$,
for various values of ${\sqrt{s}}$ and $M$, and for the 
$tu$-strength parameter $H1\!=\!0$, and in $(\ldots)$ for $H1\!=\!0.016$.
Given are also the relative increases of 
${\sigma}$ when $H1\!=\!0\!\mapsto\!0.016$.
The $N$-decay parameter $H$ is taken $H\!=\!1$; 
numbers in $[ \ldots ]$ are for $H\!=\!0.122$.}
\label{tabl1}
\end{table}

\end{document}